\documentclass[pra,twocolumn,superscriptaddress,showpacs,preprintnumbers,amsmath,amssymb]{revtex4-2}
\usepackage{graphicx}
\usepackage{dcolumn}
\usepackage{bm}
\usepackage{enumerate}
\usepackage{mathrsfs}
\usepackage{amsmath}
\usepackage{changes}
\definechangesauthor[name={Author1}, color=red]{A1}
\definechangesauthor[name={Author2}, color=green]{A2}
\usepackage{epstopdf}
\usepackage[titletoc]{appendix}
\usepackage{makecell}
\begin{document}
	
\title{Enhanced measurement precision with continuous interrogation during dynamical decoupling}

\author{Jun Zhang}
\affiliation{Key Laboratory of Artificial Micro- and Nano-structures of Ministry of Education,
	and School of Physics and Technology, Wuhan University, Wuhan, Hubei 430072, China}

\author{Peng Du}
\affiliation{Key Laboratory of Artificial Micro- and Nano-structures of Ministry of Education,
	and School of Physics and Technology, Wuhan University, Wuhan, Hubei 430072, China}

\author{Lei Jing}
\affiliation{Key Laboratory of Artificial Micro- and Nano-structures of Ministry of Education,
	and School of Physics and Technology, Wuhan University, Wuhan, Hubei 430072, China}
\author{Peng Xu}
\affiliation{Key Laboratory of Artificial Micro- and Nano-structures of Ministry of Education,
	and School of Physics and Technology, Wuhan University, Wuhan, Hubei 430072, China}

\author{Li You}
\affiliation{State Key Laboratory of Low-Dimensional Quantum Physics, Department of Physics,Tsinghua University, Beijing 100084, China}

\author{Wenxian Zhang}
\email[Corresponding email: ]{wxzhang@whu.edu.cn}
\affiliation{Key Laboratory of Artificial Micro- and Nano-structures of Ministry of Education,
	and School of Physics and Technology, Wuhan University, Wuhan, Hubei 430072, China}
\affiliation{Wuhan Institute of Quantum Technology, Wuhan, Hubei 430206, China}

\begin{abstract}
Dynamical decoupling (DD) is normally ineffective when applied to DC measurement. In its straightforward implementation, DD nulls out DC signal as well while suppressing noise.
This work proposes a phase relay method that is capable of continuously interrogating the DC signal over many DD cycles. We illustrate its efficacy when applied to measurement of a weak DC magnetic field with an atomic spinor Bose-Einstein condensate. Sensitivities approaching standard quantum limit or Heisenberg limit are potentially realizable for a coherent spin state or a squeezed spin state of 10,000 atoms respectively, while ambient laboratory level noise is suppressed by DD. Our work offers a practical approach to mitigate the limitations of DD to DC measurement and will like find other applications for resorting coherence in quantum sensing and quantum information processing research.
\end{abstract}
\maketitle
\section {Introduction}
As one of the most studied control methods, dynamical decoupling (DD) can actively suppress noise to preserve coherence of a quantum system~\cite{Viola1999PRL,Suter2016RMP}, e.g., in trapped ions~\cite{Biercuk2009ion,Kotler2011,Bohnet2016Scisence,Barrett2020,Kim2021NC}, solid-state spins ~\cite{Petta2005,Liu_2007,Paola2009PRL,Lange2010NV,Medford2012PRL,Rong2014,Simon2017scienceNV,Zopes2017NV}, superconducting quantum circuits ~\cite{Nakamura2002,Bylander2011,Pokharel2018PRL}, and ultracold atomic gases ~\cite{Eto2013,Muessel_2014PRL,Trypogeorgos2018CDD,Anderson2018CDD}. Despite of its successes, DD has found only a few usages in precision measurement of a DC signal~\cite{N_one}. While unwanted and unknown stochastic noises are suppressed by periodically refocusing a quantum state to the initial one with DD~\cite{Hirose2012,Eto2013}, a DC (or low frequency AC) signal sensed from phase interrogation by a quantum state provides no net accumulated phase. Such a dilemmatic situation is unfortunate, especially since the effect of achieving beyond standard quantum limit (SQL) precision could obviously benefit from entangled or squeezed quantum states that are otherwise easily decohered~\cite{Giovannetti2004Quantum,Esteve2008,Hamley2012,pezz2018,Kaubruegger2019PRL,Bao2020,Baamara2021PRL}.

\begin{figure}[htb]
	\includegraphics[width=1.0\linewidth]{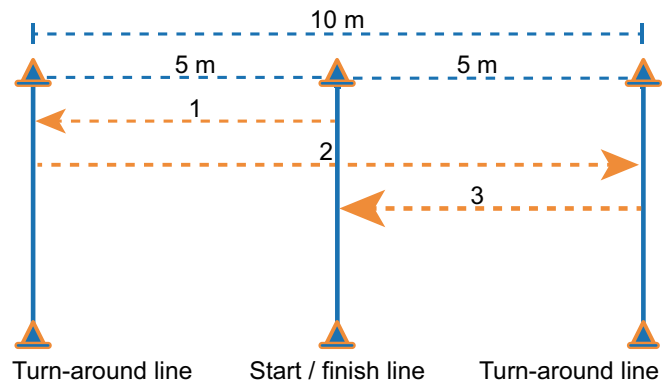}
	\caption{{Illustration of the 5-10-5 shuttle run.} The 5-10-5 shuttle run gets its name from the distance an athlete runs within each cycle, 5m to the left, 10m to the right, and 5m to the left. }
	\label{Fig.1}
\end{figure}

This work studies the application of DD to precision measurement by suppressing noise while reaping up a gain from continuous sensing of a weak DC (or low frequency AC)  signal. To explain our idea, let us recall an athlete in a 5-10-5 shuttle run.
After start, his/her displacement increases while running forward. When he/she comes around after touching the turn-around lines, however, the displacement with respect to the initial line decreases and eventually becomes zero upon returning back to the start line, although the net distance covered continuously grows after repeated line touchings at both ends (see illustration in Fig.~\ref{Fig.1}). Analogously, the interrogated signal as well as the accompanying noise accumulate during the first half of a DD cycle, but are both compensated for by the complementary second half period. To obtain a net phase accumulation of a DC signal, one could simply flip the signal's phase but not that of the noise during the second half as illustrated in Fig.~\ref{Fig.2}, provided the signal were known. Such a simple (alternately repeated) phase flip could help to recover the DC signal by allowing the consequent interrogation to last for many DD cycles while the noise remain suppressed. We refer the above described manipulation protocol as the phase relay method (PRM). If implemented this way, DD, which has been demonstrated successfully for high frequency AC signal measurements in a variety of physical systems ~\cite{Taylor2008,Kotler2011,Hirose2012,Paola2013,Eto2013,Wolf2015,Lang2015PRX,Simon2017scienceNV,Boss2017,Zopes2017NV,Degen2017}, will also be applicable to measurement of DC (and low-frequency AC) signals.

\section {Continuous interrogation during DD}

We illustrate the PRM by applying it to measure a weak DC magnetic field using an atomic spinor Bose-Einstein condensate through Ramsey interferometry~\cite{DC_ignal}. The condensate is assumed to consist of $N=10,000$  $^{87}$Rb atoms, which comes with a strong repulsive spin independent interaction in addition to a weak ferromagnetic spin exchange interaction. For small or moderate sized $N$ and with the condensate in an optical trap, its Hamiltonian under the single spatial mode approximation reduces to,
$
H = -c_2^{\prime} \bf{J}^2 + \gamma \bf{B} \cdot \bf{J},
$
after constant terms are dropped~\cite{Law98,Qze}, where $c_2^{\prime}$ denotes the spin exchange interaction strength with the collective condensate spin $\bf{J}=\sum\limits_{i=1}^{N} \bf{F}_i$ from spin $\bf{F}_i$ for $i$th atom, $\gamma$ the atomic gyromagnetic ratio, and $\bf{B}$ the total magnetic field which possibly includes a fixed known bias field $\bf{B}_0$, a fixed weak but unknown DC magnetic field $\bf{b}_0$ (the signal to be measured), and stray stochastic magnetic field $\bf{b}$ (noise) to be suppressed by DD. We consider the simple case of the bias and the signal field along the same $z$-axis direction, while the stray field is randomly directed with random amplitude ($\mid\bf{b}\mid$$< b_c$ upto a cutoff amplitude $b_c$.).

\subsection{Noise suppression with uniaxial DD}~\label{sec.S2}

The condensate spin decoheres by stray magnetic field, which eventually limits the optimal achievable sensitivity of the magnetometer. Although the dephasing noise model may give analytical results as derived in Append.~\ref{sec.FID}, we consider a more realistic stray-field model. Including the stray magnetic field, the Hamiltonian becomes
\begin{equation}
H = -c_2^\prime{\mathbf J}^2 + \gamma (B_0+b_0) J_z+\gamma (b_xJ_x+b_yJ_y+b_zJ_z)
\label{eq:hn}
\end{equation}
where $b_{x,y,z} \in [-b_c, b_c]$ and a bias field $B_0\gg b_{0,c}$ is required. To suppress decoherence from stray field, many DD schemes can be applied~\cite{Hahn1950Spin,Meiboom1958Compensation,solomon1959rotary, Viola1999Dynamical,Uhrig2007, DasDarma2008,Uhrig2009Concatenated,Haeberlen2012High,Suter2016RMP,Jun2016}. We adopt a uniaxial DD protocol which is robust and easy to implement experimentally~\cite{Yao2019UniDD}.

By denoting the $\pi$-pulse along  $x$-axis by $X$, the evolution operator for a unit cycle becomes $U_{2\tau}=XU_{\tau}XU_{\tau} = \exp{(-i2\tau \bar{H})}$, for $U_\tau=\exp(-i\tau H)$ with $\tau$ the duration between pulses and $\bar{H}$ the time averaged Hamiltonian over a whole cycle. Under the magic condition $\tau=2m\pi/(\gamma B_0)$ (for $m=1,2,3,...$), the uniaxial DD is capable of efficiently suppressing decoherence from stray magnetic field, because the effective coupling between the condensate spin and the stray magnetic field is reduced by a factor of $(b_z/B_0)$, leading to $\bar{H}\approx -c_2^\prime{\mathbf J}^2+ (b_z/B_0) \gamma b_yJ_y$~\cite{Yao2019UniDD}. Further suppression of the stray magnetic field is achieved if a refined uniaxial DD protocol, the balanced Uniaxial DD (BUni-DD) $[X\bar{U}_{\tau}X\bar{U}_{\tau}XU_{\tau}XU_{\tau}]$, is employed, where $U_{\tau}$ ( $\bar{U}_{\tau}$) denotes free evolution with a positive (negative) bias field. The latter leads to the BUni-DD period $T=4\tau$ and the magic condition $\gamma B_0\tau=2m\pi$ remains needed. It is easy to find the the average Hamiltonian further improves to $\bar{H}\approx -c_2^{\prime}{\mathbf J}^2+O((b_c/B_0)^2)$.

\subsection{Phase relay method}~\label{sec.S3}

\begin{figure}[tb]
	\includegraphics[width=1.0\linewidth]{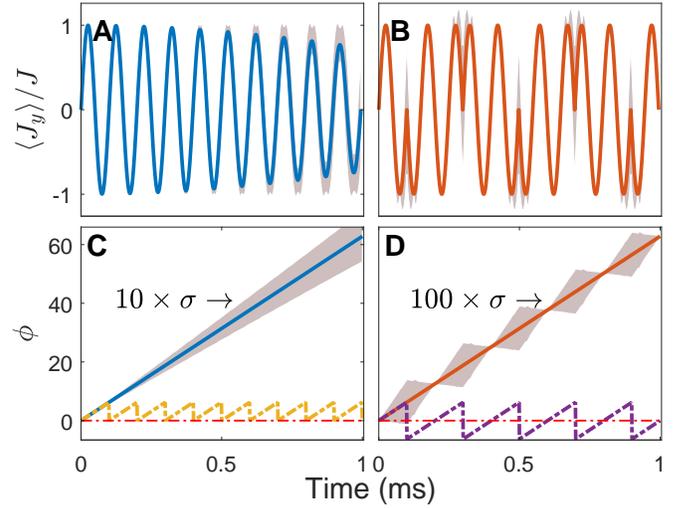}
	\caption{{Phase relay method.} ({A}) Numerically simulated mean (blue solid line) and the standard deviation (gray shaded region) for observable $J_y$ without DD. ({B}) The same as in (A) but with DD shown
	by the red solid line (mean) and the gray shaded region (one standard deviation). The applied pulse interval is $\tau = 0.1$ ms. The standard deviation is found to be strongly suppressed after one DD cycle $T=4\tau$. ({C}) Phase $\phi$ extracted directly from the simulation (A) (yellow dash-dotted line) and with hypothetical accumulation taking into account of the indexed number of oscillations (blue solid line). The standard deviation of the phase (shaded in gray) is multiplied by 10 for better viewing. The phase increases linearly and the standard deviation grows monotonically with time. ({D}) Phase extracted directly from the experimental observable (B) (purple dash-dotted line) and after applying the PRM (red solid line). The standard deviation of the phase (shaded gray) is multiplied by 100. Similarly, the phase after PRM increases linearly and the standard deviation is strongly suppressed after each DD cycle.}
	\label{Fig.2}
\end{figure}

To estimate the signal field amplitude with Ramsey interferometry, the condensate spin is initially prepared to align along $x$-direction. During the interrogation time $t$, it precesses along the direction of the bias and signal field sum which provides an accumulated time phase $\phi = \gamma (B_0+b_0) t$ with $B_0=\mid\bf{B_0}\mid$ and $b_0=\mid\bf{b_0}\mid$. By measuring precisely this phase $\phi$ at time $t$, the signal field can be determined. Experimentally, this is accomplished by measuring $J_y$,  the $y$-component of the collective spin, which oscillates with time as shown in Fig.~\ref{Fig.2}. The estimated phase between $0$ and $2\pi$ is extracted by fitting the observed data to a sinusoidal function. For longer interrogation times, this phase will exceed $2\pi$. Consequently, one would hope to keep track of indexing the number of oscillation cycles in order to continuously interrogate a monotonically increased phase instead of a zigzaged one chopped into $[0, 2\pi]$. During interrogation, the stray magnetic field stochastically disturbs the measured phase and introduces an uncertainty $\Delta \phi$, which leads to an uncertainty in the signal $\Delta b_0$. Such effects from stray magnetic field can be suppressed by periodically executed DD pulses, for example with instantaneous $\pi$ rotations of the spin along $x$-axis, or spin flips. As numerical results verify in Fig.~\ref{Fig.2}, effects of the noise are indeed suppressed significantly. Unfortunately, the interrogated phase from the bias plus the signal field is also refocused by the above DD pulses, i.e., the net phase accumulation after complete cycles of DD diminishes, making it difficult to estimate the signal field.

\begin{figure}
	\centering
	\includegraphics[width=1.0\columnwidth]{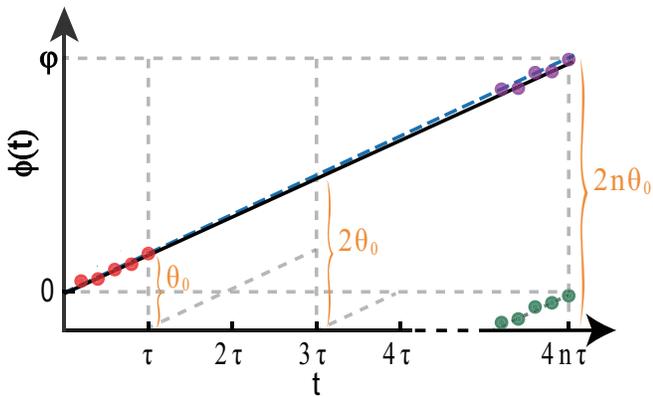}
	\caption{The PRM protocol. A roughly estimated phase-shift $\theta_0$ is obtained by linear least square fitting (red dots) to a few data points extracted from  preliminary measurements of the observable during the first quarter period. One then extracts an estimated phases $\phi(t)$ (green dots) and shifts them according to $\theta_0$ and the cycle indexing number $n$ (purple dots), during the whole BUni-DD sequence. The blue dashed line illustrates the relayed phase based on the preliminary estimated $\theta_0$. An improved estimate for $\theta_0$ is obtained by a least square linear fit to all the relayed data from many BUni-DD cycles (black line). The slope of the fitted black line denotes an refined estimate of the signal field $b_0$ .}
	\label{Fig.3}
\end{figure}

While BUni-DD protocol discussed above suppresses stray magnetic field, the signal field is also suppressed unfortunately as the accumulated phase nearly zeros out after each DD cycle. However, the signal field differs from stray magnetic field because it is by definition nonzero and we further note the pulse interval for the DD pulse sequence $\tau$ is exactly known. These features (knowledge) provide us an opportunity to continuously accumulate the phase associated with the signal by the phase relay method (PRM), inspired by the 5-10-5 shuttle run. 

We now examine the interrogated phase accumulation associated with the signal in the first DD cycle, over a period of $4\tau$ (see Fig.~\ref{Fig.3}). For the sake of clarity, we neglect stray magnetic field at this stage. During the first quarter, the phase of the signal $\varphi(\tau) = \gamma b_0 \tau$ accumulates linearly. At $t=\tau$, an instantaneous $X$ pulse then flips the condensate spin and the phase changes to $-\varphi(\tau)$. During the second quarter, the signal's phase evolves linearly back to $0$, i.e., $\varphi(2\tau)=0$. 
This phase pattern repeats in the third and the fourth quarters as in the first and the second quarters, respectively. The total interrogated phase during a whole DD cycle vanishes.

As shown in Fig.~\ref{Fig.3}, an immediate method to continuously interrogate the signal phase during the BUni-DD cycle is to shift the phase during $t\in[\tau, 3\tau]$ by a fixed amount $\theta_0$  and by $2\theta_0$ during the last quarter $t\in[3\tau, 4\tau]$ respectively, provided $\theta_0 = \gamma b_0 \tau$ from the signal field $b_0$ were known. For the $n$th DD cycle, the required phase shift of the first quarter becomes $(2n-2)\theta_0$, the second and the third quarter $(2n-1)\theta_0$, and the fourth quarter $2n\theta_0$. In practice, we would first obtain a crude estimate to the phase shift $\theta_0$ by running a few preliminary experiments  without DD pulses, then followed by optimizations, and finally determine $\theta_0$ more precisely. One plausible approach is shown in Fig.~\ref{Fig.3} which obtains the shifted phase data (by varying $\theta_0$) during the first and the last quarters of the whole DD sequence, and optimizes with least square fitting. Afterwards, a refined and more accurate $\theta_0$ may be determined and likewise the signal field is determined more precisely.

Including stray magnetic field, our numerical simulations confirm the PRM remains effective. For practical implementations, the PRM protocol can be summarized explicitly by the following steps.
\begin{itemize}
	\item[(i)] Take a few measurements of $\langle J_y\rangle$ to obtain $\langle J_y\rangle(t)$ at short times.
	\item[(ii)] Extract the phase $\varphi(t)$ from the experiment data, fit linearly the phase-time curve, and find out a crude estimate for the phase-shift $\theta_0$, which is similar to the adjustment of fire in military.
	\item[(iii)] Take more measurements of $\langle J_y\rangle(t)$ at the beginning, in the middle, and at the end of the BUni-DD sequence.
	\item[(iv)] Extract phases $\varphi(t)$ from the larger data set, and relay/shift measured phases accordingly with the knowledge of the cycle indexing number $n$ and $\theta_0$.
	\item[(v)] Optimize $\theta_0$ with least-square linear fit.
	\item[(vi)] Derive the signal field $b_0$.
\end{itemize}

With the PRM, the phase relays continuously during a DD cycle, as shown schematically in Fig.~\ref{Fig.2}, which can be further optimized by least square fitting the data over many DD cycles. Consequently, the net phase from the bias and signal fields continuously integrates while the stochastic contributions from noise are averaged out after repeated DD cycles, thus the measurement precision of a weak DC signal is enhanced.

\section{Heisenberg limit magnetometer with spin squeezing}

\begin{figure*}[tb]
	\includegraphics[width=0.8\linewidth,scale=1.00]{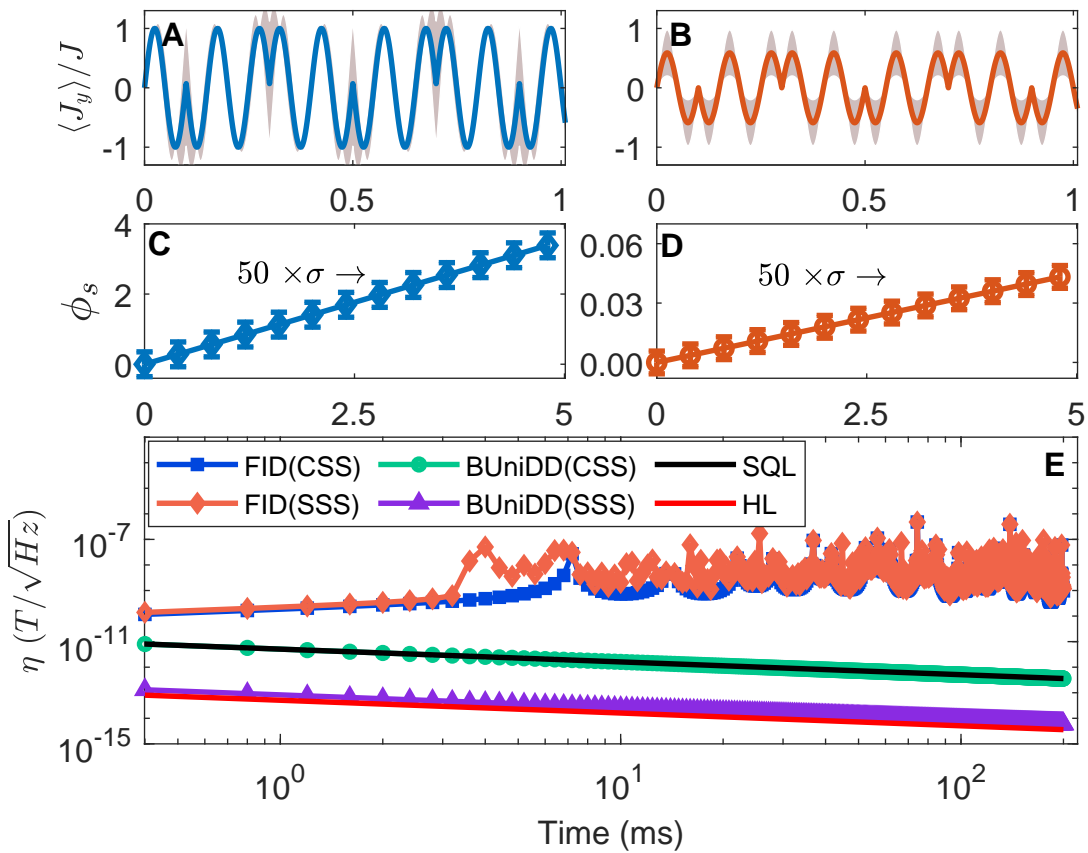}
	\centering
	\caption{{Precision measurement of a weak DC magnetic field with DD.} ({A, B}) The average condensate spins (solid lines) and their standard deviations (gray shaded areas) for initial coherent spin state (CSS) versus squeezed spin state (SSS) ({A, B}).  ({C, D}) The signal's phases $\phi_s$ extracted with PRM from simulated experimental observables in ({A}) and ({B}), respectively in ({C}) and ({D}) with  $b_0 = 160$ $\mu G$ in ({C}) and $b_0 = 1.6$ $\mu G$ in ({D}). The bias and the cutoff fields are assumed to be $B_0 = 14.3$ mG and $b_c = 0.1$ mG, respectively. The small standard deviations are multiplied by 50 in order to be clearly visible. ({E}) The signal field's sensitivities under DD for CSS (green circles) and SSS (purple triangles), respectively approaching the SQL (black solid line) and HL (red solid line). For comparison, we present the sensitivities without DD pulses for both CSS (blue squares) and  SSS (red diamonds) as well in the same figure. The DD pulses are seen to strongly suppress noise and significantly improve sensitivity. }
	\label{Fig.4}
\end{figure*}

To evaluate the achievable precision after an interrogation time $t$, we calculate the signal-field sensitivity for measuring the weak magnetic field according to
\begin{equation*}
\eta = \frac{\Delta J_y}{\mid\partial J_y /\partial \phi\mid \gamma \sqrt{t} }\;,
\end{equation*}
which usually diverges at very short time as stochastic uncertainty from the stray field is yet to become effective~\cite{Taylor2008,Degen2017,Boss2017}.
In order to achieve quantum enhanced sensitivity, it is necessary to consider metrologically useful entangled states such as the spin squeezed states (SSS)~\cite{pezz2018}. In the absence of stray field, an SSS is capable of approaching the HL scaling, i.e. $\eta_{HL}< \eta <2\eta_{HL}$, where $\eta_{HL}={1}/{(\gamma J\sqrt{t})}$ (see Append.~\ref{sec.S1} for detail derivation)~\cite{Giovannetti2004Quantum,Esteve2008}.

We consider an SSS with averaged spin point along the $x$-axis, while the squeezing direction is along the $y$-axis, orthogonal to the averaged spin. Such a state is readily prepared experimentally in a spinor $^{87}$Rb atomic condensate~\cite{Hamley2012,Luo2017,Zou2018}. We set $\omega_0 = \gamma B_0 = 2\pi \times 10$ KHz corresponding to a bias field of $B_0 = 14.3$ mG ($\gamma = 0.70$ MHz/G). The cutoff field is taken to be at a nominal laboratory noise level $b_c = 0.1$ mG (with an angular frequency $2\pi \times 70$ Hz). As for the signal, we assume $b_0 =1.6$ $\mu$G, which serves to demonstrate quantum enhanced precision. The pulse interval of Buni-DD is $\tau = 0.1$ ms, satisfying the magic condition $\omega_0\,\tau = 2 \pi$. After repeated simulation runs with random stray magnetic field $\bf{b}$ of uniformly distributed components $b_{x,y,z}\in [-b_c, b_c]$, we compute the observable $\langle J_y\rangle$ and its standard deviation, extract the phase, derive the signal's phase and its standard deviation, and finally determine the sensitivity $\eta$ (Fig.~\ref{Fig.4}).

\begin{figure}
	\centering
	\includegraphics[width=1.0\columnwidth]{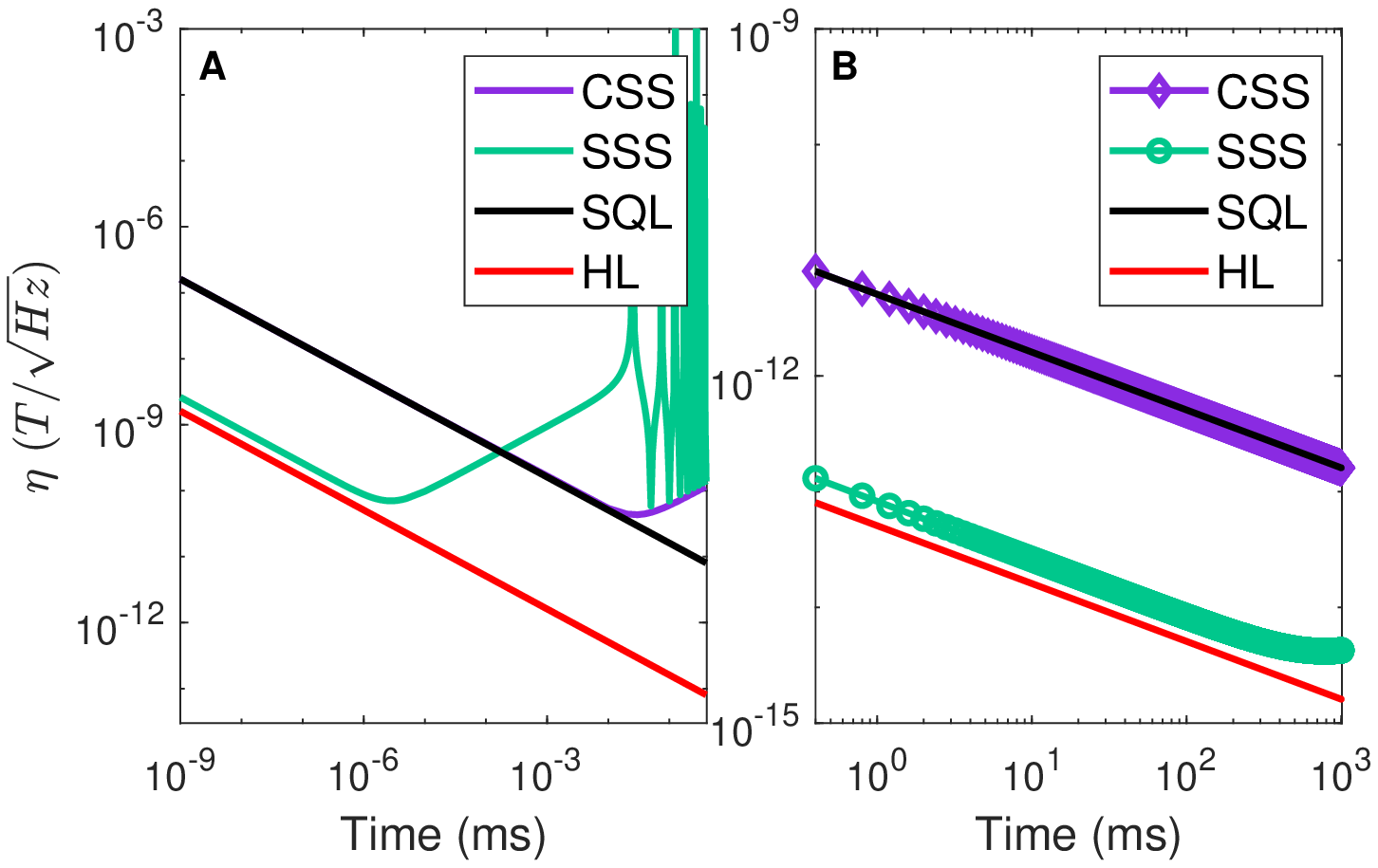}
	\caption{Sensitivity at short and long times for CSS (purple line) and SSS (green line). The cutoff field assumed $b_c=0.1$ mG. Both states decohere quickly without the application of BUni-DD ({\bf{A}}) but maintain their coherence for longer times with the application of BUni-DD ({\bf{B}}). The black and the red lines denote SQL and HL, respectively.}
	\label{Fig.5}
\end{figure}

Although an entangled SSS can enhance sensing sensitivity, in practice this is difficult because an entangled state is fragile to stray magnetic field.  Figure~\ref{Fig.5}(A) shows the sensitivity under free evolution within $1$ ms, which decoheres quickly such that no quantum enhancement is observed. The oscillations of sensitivity arise due to the periodicity in Eq.~(A.7) and the optimal value is found to be limited by $\sim 100$ $pT/\sqrt{Hz}$. Therefore, it is crucial to efficiently suppress stray magnetic field for sensing a weak signal field, especially when entangled probe states are used.

By applying BUni-DD pulses and performing the PRM to the SSS, we obtain much higher sensitivity going beyond SQL and approaching HL within $200$ ms, at about 6 fT/$\sqrt{Hz}$ for 10,000 atoms. The projected sensitivity continuously improves with prolonged interrogation time. As shown in Fig.~\ref{Fig.5}(B), the sensitivity for the SSS reaches its optimum $\sim 4$ $fT/\sqrt{Hz}$ at $t\sim 770$ ms. The simulation results show clear quantum enhancement.

\begin{figure*}[tbh]
	\includegraphics[width=0.8\textwidth]{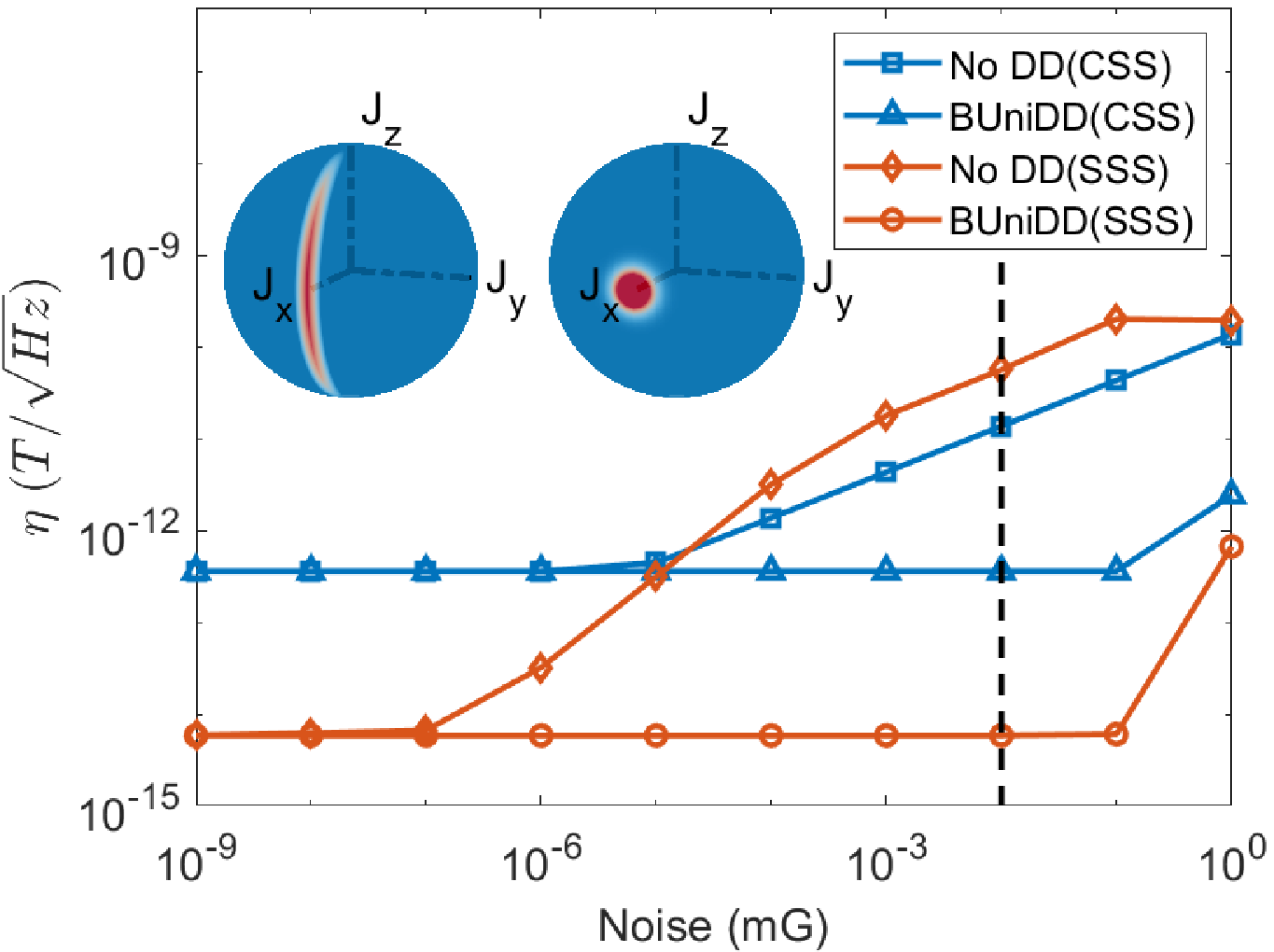}
	\caption{{Optimal sensitivities within $200$ ms of interrogation time at different noise levels}. The optimal sensitivities detected by an initial CSS with (without) DD pulses are denoted by blue triangles (squares) versus the ones detected by an initial SSS with (without) DD pulses are shown with orange circles (diamonds). As long as stray magnetic field is lower than the noise threshold $\sim 0.1$ mG, noise is strongly suppressed and sensitivities with DD are independent of the noise strength. For comparison, the noise thresholds without DD pulses are numerically found to be $\sim 10^{-6}$ mG for an initial CSS and $\sim 10^{-8}$ mG for an initial SSS. The vertical line marks a laboratory noise level of $0.01$ mG~\cite{Eto2013}.}
	\label{Fig.6}
\end{figure*}

To investigate the dependence of sensitivity on the noise field strength, we change the cutoff field $b_c$ from $10^{-9}$ mG to $1$ mG, but fix the interrogation time at $t=200$ ms while keeping other parameters the same as in Fig.~\ref{Fig.4}. The optimal sensitivity at low noise level is found to settle at about 6 fT/$\sqrt{Hz}$ and is independent of $b_c$ (Fig.~\ref{Fig.6}). The sensitivity deteriorates rapidly with increasing noise strength once $b_c$ is larger than a threshold, which is found to be approximately $1/(\gamma t J)\sim 10^{-7}$ mG without DD pulses but becomes $(B_0/b_c)^2/(\gamma t\sqrt J)\sim 0.1$ mG with DD pulses. Although the SSS is fragile to stray field, the DD implemented with our PRM supports its application to precision measurement at common laboratory noise level ($b_c \sim$ 0.1 mG)~\cite{Eto2013,Gross2010nonlinear}.

As for an initial quantum state without entanglement between atoms, such as coherent spin state (CSS), the optimal sensitivity follows the SQL in the absence of stray magnetic field. Similarly, the CSS decoheres within $0.1$ ms in the presence of stray field, as shown in Fig.~\ref{Fig.5}(A). However, after the application of BUni-DD with PRM, the sensitivity of CSS is found to reach SQL in the same noisy environment. Assuming the same number of atoms and averaged spin direction as the SSS, we achieve the sensitivity 0.36 pT/$\sqrt{Hz}$ for sensing a signal field $b_0 = 160$ $\mu$G. This result compares favorably to a previously reported experimental result of $0.5$ pT/$\sqrt{Hz}$ with $N\sim 10^6$ atoms for a unity duty cycle~\cite{stamper2007magnetometry}. As shown in Fig.~\ref{Fig.5}(B), the sensitivity for CSS keeps tracking SQL within an interrogation time of 1 s, due to application of BUni-DD and PRM. In addition, the threshold of noise strength that impairs the sensitivity of a CSS is approximately $1/(\gamma t\sqrt J)\sim 10^{-5}$ mG without DD pulses, but lifts up to $(B_0/b_c)^2/(\gamma t\sqrt J)\sim 0.1$ mG with DD pulses.



\section{Conclusions and discussions}

In conclusion, we propose the PRM which enables continuous interrogation during DD and enhances the system's coherence during quantum measurements. With this PRM, we numerically show that the sensitivity of a spinor-BEC-based magnetometer approaches the SQL with an initial CSS and the HL with an SSS. Remarkably, the simulated optimal sensitivity with only 10,000 atoms enters the regime of a few fT/$\sqrt{Hz}$ within 200 ms interrogation time, indicating the great potential of enhanced quantum metrology utilizing the spin squeezing and the metrologically-useful-quantum entanglement. The PRM applies not only to the investigated spinor-BEC magnetometer~\cite{Eto2013,Muessel_2014PRL}, but also to other quantum sensor systems suffering severe decoherence, such as Nitrogen vacancy centers in diamond~\cite{Petta2005,Liu_2007,Paola2009PRL,Medford2012PRL,Rong2014,Zopes2017NV}, trapped ions~\cite{Kotler2011,Bohnet2016Scisence,Barrett2020} and superconducting quantum devices~\cite{Nakamura2002,Bylander2011}.

Three concerns are experimentally related. (i) At a first glance, the relative stability of the bias field $b_0/B_0$ is roughly 0.01\%. Although such a highly stable magnetic field has been realized in recent experiments, where the bias magnetic field is of the order of $\sim 1\; G$  and stabilized to a range from $10\; \mu G$ to $0.1$ mG~\cite{Kim2021NC,Luo2017,Zou2018}, we note that the magnetic field stability is actually characterized by the stray magnetic field. Thus the ratio $b_c/B_0$ is about 1\% which is quite easy to realize. (ii) The rapid reversion of the bias field (in the order of $\sim 1 \,\mu s$) was achieved in experiments without introducing adiabatic Landau-Zener transitions between atoms in different spin components~\cite{Kell2021}. Therefore we may neglect the short reversion time of the bias field during our numerical simulations. (iii) To measure $\langle J_y\rangle$ in a spin-1 BEC experiment, one may employ continuous measurement using magneto-optical imaging techniques~\cite{stamper2007magnetometry}, or projective measurement using the Stern-Gerlach technique after an additional $\pi/2$ pulse which rotates the condensate spin to $z$ direction~\cite{Luo2017}. 

For the purpose of a concise illustration of PRM, we have assumed several idealizations for the magnetic field sensitivities to reach the HL. Specifically, the sensitivity calculations do not take account of any dead time during state-preparation and measurement stage, and assume hard pulses with instantaneous and perfect rotation. Therefore, the sensitivities achieved in practical experiments are not as prominent as predicted here. Nevertheless, our results point out the possibility to suppress noises with DD for a DC magnetometer using PRM. It is worthwhile to design more robust pulse sequences accounting for pulse width and shape, in order to meet practical requirements.

\section{Acknowledgments}

This work is supported by the NSAF (grant no. U1930201), the Natural Science Foundation of China (grant nos. 12274331, 91836101, and 91836302), the National Key R\&D Program of China (grant no. 2018YFA0306504), and Innovation Program for Quantum Science and Technology (grant no. 2021ZD0302100). The numerical calculations in this paper have been partially done on the supercomputing system in the Supercomputing Center of Wuhan University.

\begin{appendix}
	

	

	
	\section{Parameter estimation and sensitivity}~\label{sec.S1}
	
	The Hamiltonian for the magnetometer based on atomic spin-1 Bose-Einstein condensate is,  under the single mode approximation~\cite{Law98}, given by,
	\begin{eqnarray} \label{eq.2}
		H = -c_2^\prime {\mathbf J}^2 + \gamma {\mathbf B}\cdot {\mathbf J}
	\end{eqnarray}
	where $c_2^\prime$ represents the spin exchange interaction strength,  $\bf{J}=\sum\limits_{i=1}^{N} \bf{s}_i$ the collective atomic spin of the condensate, and $\bf{s}_i$ the spin operator for the $i$th atom, $\gamma=(2\pi)\; 0.70$ MHz/G the gyromagnetic ratio, and ${\mathbf{B}}$ the total magnetic field which includes  a DC magnetic field $b_0$ (signal),   stray magnetic field $b$ (noise), and possibly a bias $B_0$. The direction and magnitude of the DC field are fixed, while they are random for the stray field.
	
	With the (to be estimated) signal magnetic field $b_0$ assumed static and pointed along the $z$-axis, the Hamiltonian in the absence of stray field reduces to, $H_0 = -c_2^\prime {\mathbf J}^2 + \gamma b_0 J_z$. To estimate $b_0$, a polarized initial state $\psi(0)$, e.g. a coherent spin state (CSS) with its average spin along $x$-direction taking the maximum collective spin value $\langle J_x\rangle=J$ is prepared. It evolves under $H_0$ for an interrogation time $t$ and becomes $\psi(t)=\exp{(-i[-c_2^{\prime }t\mathbf{J}^2+\varphi J_z])}\psi(0)$, where $\varphi=\gamma b_0 t$ is the accumulated dynamical phase from the signal field $b_0$. One can determine $b_0$ by measuring a dependent observable, for instance $\langle J_y\rangle$, which gives $\langle J_y\rangle(t)=J\sin{(\varphi)}$. According to  parameter estimation theory, the value of $b_0$ can be estimated from the measured $\langle J_y\rangle(t)$, through the relation
	\begin{eqnarray}
		b_0 &=&\frac{1}{\gamma t}\arcsin{\left(\frac{\langle J_y\rangle(t)}{J}\right)}.
	\end{eqnarray}
	
	The variance of the measurement data limits the uncertainty of $b_0$. The minimal detectable magnetic field for an interrogation time $t$ is thus found to be
	\begin{eqnarray}
		\Delta b_0 &=&\frac{1}{\gamma t}\frac{\Delta J_y}{\mid\partial{\langle J_y\rangle(t)}/\partial{\varphi}\mid},
	\end{eqnarray}\label{eq.DeltaB}
	where $\Delta J_y=\sqrt{\langle J^2_y\rangle-\langle J_y\rangle^2}$ denotes the standard
	deviation. When $M$ uncorrelated consecutive measurements are carried out for a total measurement time $T=Mt$, the minimal detectable field becomes
	\begin{eqnarray}
		\Delta b_0 &=& \frac{1}{\gamma}\frac{1}{\sqrt{t T}}\frac{\Delta J_y}{\mid\partial{\langle J_y\rangle(t)}/\partial{\varphi}\mid}.
	\end{eqnarray}
	
	The sensitivity for such a magnetometer is defined as
	\begin{eqnarray}
		\eta&=&\Delta b_0\sqrt{T}\nonumber\\
		&=&\frac{1}{\gamma}\frac{1}{\sqrt{t}}\frac{\Delta J_y}{\mid\partial{\langle J_y\rangle}/\partial{\varphi}\mid}.
	\end{eqnarray}
	It is straightforward to find that the optimal sensitivity for an initial CSS is given by
	\begin{eqnarray}
		\eta_{C}(t)=\frac{1}{\gamma}\frac{1}{\sqrt{2t J}},
	\end{eqnarray}
	which reaches the standard quantum limit (SQL) with respect to the spin size $J$ following the scaling $1/\sqrt{J}$.
	
	The sensitivity for a squeezed spin state (SSS) with average spin pointed along the same $x$-direction exhibiting optimal squeezing along $y$-direction is given by
	\begin{eqnarray}
		\eta_{S}(t)=\frac{1}{\gamma}\frac{\theta(t)}{\lambda}\frac{\sqrt{\xi^2_s}}{\sqrt{2t J}},
	\end{eqnarray} \label{eq.eta_sss}
	
	where $\theta(t)=\sqrt{1+\tan^2{(\omega t)}(\langle\Delta^2 J_{x}\rangle_{0}/\langle \Delta^2 J_{y}\rangle_{0})}$ corresponds to the fixed direction measurement (e.g. $y$-direction) and $\langle\cdots\rangle_0$ denotes the average over the initial state. It is important to point out the optimal squeezing direction remains perpendicular to the averaged spin for such a SSS during the Larmor precession in a magnetic field along $z$-axis. Nevertheless, the optimal squeezing direction is difficult to track experimentally, so we measure spin squeezing in the fixed $y$-direction. $\lambda=\mid\langle J_x\rangle_0\mid/J$ represents the ratio of the average initial spin to the total spin size, with $\lambda=1$ for a CSS and $\lambda\approx 0.58$ for the SSS at $J=10,000$ respectively. Following Kitagawa and Ueda ~\cite{Kitagawa1993}, we define spin squeezing parameter $\xi^2_s=2 \min(\Delta^2 J_y)/J$, which gives $\xi^2_s=1$ for a CSS and $\xi^2_s\approx 0.91/J$ for the SSS at $J=10,000$ in a spin-1 atomic Bose-Einstein condensate (as considered in the main text). At $\theta(t)=1$, we find $$\eta_{S}\approx \frac{1.6}{\gamma}\frac{1}{\sqrt{2t}\;J},$$
	an optimal sensitivity that scales as $1/J$,
	which is beyond SQL and approaches to the Heisenberg limit (HL). The scaling with respect to $J$ manifests the quantum enhancement to measurement precision by using SSS.

\section{Free evolution in dephasing noises}\label{sec.FID}

To facilitate analytical derivation, we consider the dephsing model in which the stray field distributed uniformly within $[-b_c,b_c]$ only along $z$-direction. The system Hamiltonian becomes $H = -c_2^{\prime} {\mathbf J}^2 + \gamma b_z J_z + \gamma b_0 J_z$, where $b_z$ denotes the $z$-component of the stray magnetic field.

It is easy to calculate straightforwardly the evolution of $\langle J_y\rangle$ for an initial CSS with spin initially polarized along $x$-direction, which follows,
\begin{eqnarray}
	\label{eq.fid}
	\langle J_y \rangle(t)=J\sin{(\omega_s t)}\frac {\sin{(\omega_ct)}}{\omega_ct}.
\end{eqnarray}
with $\omega_{s/c}=\gamma b_{0/c}$. The cutoff Larmor frequency $\omega_c$ characterizes a dephasing time $T^{\ast}_2\sim \omega^{-1}_c$. The variance $(\Delta J_y)^2$ can be divided into a quantum and a classical part,
\begin{eqnarray}
\label{eq.B2}
	(\Delta J_y)^2(t)=(\Delta J_y)^2_Q(t)+(\Delta J_y)^2_C(t),
\end{eqnarray}
where
\begin{eqnarray}
(\Delta J_y)^2_Q(t) &=& {J\over 4}\left[1+ \cos{(2\omega_st)}{\sin{(2\omega_ct)}\over {2\omega_ct}}\right], \\
(\Delta J_y)^2_C(t) &=& {J^2\over 2} \left[1-\cos{(2\omega_st)}{\sin{(2\omega_ct)}\over {2\omega_ct}}\right. \nonumber\\
&&\left.-2\sin^2{(\omega_st)}{\sin^2{(\omega_ct)}\over {(\omega_c t)^2}}\right].
\end{eqnarray}
At ultrashort time when $\omega_c t\ll 1$, only the quantum part plays a role because  $(\Delta J_y)^2_Q(t)\approx ({J}/{4})[1+\cos{(2\omega_st)}]$ while $(\Delta J_y)^2_C(t)\approx 0$. As time goes by, the classical part becomes more and more important and eventually dominates at long time for large $J$. Similar to the results shown in Fig.~\ref{Fig.2}(A), $\langle J_y\rangle(t)$ exhibits decayed oscillations, while the standard deviation $\Delta J_y$ increases with time, according to Eqs.~(\ref{eq.fid}) and~(\ref{eq.B2}).

\end{appendix}
%



\begin{thebibliography}{56}
\bibitem [{DC_()}]{DC_ignal}%
	\BibitemOpen
	\href@noop {} {}\bibinfo {note} {DC signal here refers to a static (or low
		frequency) magnetic field to be measured, while an observable such as spin
		average is oscillating with a (high) Larmor frequency.}\BibitemShut {Stop}%
\bibitem [{Qze()}]{Qze}%
	\BibitemOpen
	\href@noop {} {}\bibinfo {note} {The quadratic Zeeman shift for $^{87}$Rb
		scales as 72 Hz/G$^2$, resulting negligible quadratic energy for our
		calculation at a typical magnetic field ($\sim 10$ mG).}\BibitemShut {Stop}%
\end{thebibliography}



\end{document}